\documentstyle[12pt]{article}
\def\A{\widehat{Area}}
\begin{document}
\title{A 2-Surface Quantization of the Lorentzian Gravity}

\author{Marcin Bobie\'nski, Jerzy Lewandowski and Mariusz Mroczek}
\maketitle

\centerline{\it Instytut Fizyki Teoretycznej, Uniwersytet Warszawski,
ul. Hoza 69, 00-629 Warszawa, Poland}

\begin{abstract} This is a contribution to the MG9 session QG1-a.
A new quantum representation for the
Lorentzian gravity is created from the Pullin vaccum by the operators assigned
to 2-complexes. The representation uses the original, spinorial
Ashtekar variables, the reality conditions are well posed
and Thiemann's Hamiltonian is well defined. The results
on the existence of a suitable Hilbert product are partial.
They were derived in collaboration with Abhay Ashtekar.
\end{abstract}

The canonical gravity in terms of the Ashtekar variables
is often compared to the complex representation of the harmonic
oscillator. Indeed, given a 3-dimensional space of the Cauchy data
the Ashtekar connection \cite{A} is  $\Gamma + iK$ where, in the language
of the standard ADM variables,  $\Gamma$ is the Riemann connection
of a 3-metric tensor defined on $M$, and $K$ represents
the extrinsic curvature. The  variable canonically conjugate
to $A$ is an sl(2,C)${}^*$ valued  2-form $E$. Its relation
with the 3-metric can be given by the 2-area of the parallelogram
formed by $X,Y\in T(M)$:
$Area^2(E)_{(XY)}=-\frac{1}{2}{\rm Tr}(E_{XY}E_{XY})$. The Poisson
bracket is $\{A^i_a(x), E_{jbc}(y)\}= i\epsilon_{abc}\delta^i_j
\delta(x,y)$. The reality conditions of a pair $(A,E)$ can be
formulated in the following way: $(i)$ $E$  defines a
real, non-zero  area for every pair $X,Y$ of independent vectors,
$(ii)$ $dE^i + \frac{1}{2}\epsilon_{ijk}A^j\wedge E^k = 0$,
and $(iii)$ the multiple  Poisson brackets of the areas with
the Hamiltonian, that is $\{...\{Area(E)_{(XY)},
H\}...,H\}$ are all real.
Those conditions imply the relation $A=\Gamma(E)+iK$ invoked at the beginning,
the relation  is not assumed in the reality conditions.

By analogy with the harmonic oscillator and due to the dualities
connections-curves and 2-forms-2-surfaces one may consider
the functions  $A \mapsto {\rm Tr\ Pexp}\int_p-A$ defined by the
holomy along loops $p$ in $M$ as the quantum states. However,  we are motivated
by the dual representation of the quantum states
of the oscillator. Then, the vacuum state  $< 0|_{\rm osc}$ is the Dirac delta
at `position $+$ $i$ momentu', whereas the excited states are the
derivatives of
the delta distribution. Our states are taken to be linear functionals defined
on  the algebra $\rm Cyl$ of functions of $A$  generated by all the
loop functions $\psi(A) =  {\rm Tr\ Pexp}\int_p-A$.
For the  vacuum state $< 0|$ of the
Ashtekar-Einstein gravity we take the linear functional defined
 by $< 0|:\psi\ \rightarrow \psi(A=0)$. This state was first observed
by Pullin.
Notice, that the classical point $A=0$ corresponds to the  data
representing the flat spacetime, and the state is not in any sense
trivial. `Excited' states are `created'
by the dual action of suitably selected operators available
in ${\rm Cyl}$. In that way we defined a certain vector space ${\cal H}$ which
contains  the states given by the algebra of operators
generated by all the area operators \cite{AL} and all the holonomy operators,
the last ones acting by multiplication.  An element of this space
can be labelled by a 2-complex contained in $M$, and
sets of holomorphic functions $f: sl(2,C)^m\rightarrow C$
assigned to each site, edge and vertex of the complex.
An important property of ${\cal H}$ is that Thiemann's regularisation
\cite{Thomas}
of the Hamiltonian operator can be applied therein,
and the space is preserved by the Hamiltonian provided
we use a constant laps function.

The choice of a diffeomorphism invariant Hilbert product in
${\cal H}$ is an open issue. The reality conditions come down to  the
self-adjointness
of the area operators and the Hamiltonian operator.
We studied independently two samples. The self-adjointness issue of
the area operator $\A_S$ was  studied in the  restriction to the subspace
of ${\cal
  H}$ given by the surface $S$. The self-adjointness can be
achieved by using a measure on sl(2,C) supported on a region
of the algebra such that the function $f(w) := -{\rm Tr} w^2$
is positive definite. The  dependence of a Hilbert product on the
complexes, was examined in the subspace consisting of the states
$<0|\A_S$ given by all the area operators.
Therein, we have constructed all the bilinear forms that are
topologically invariant. (Recall that this symmetry is a feature
of the natural integral defined in the algebra ${\rm Cyl}$ in the
SU(2) group case). It turns out, that a 2-torus has the zero
norm with respect to each of them.  Therefore a diffeomorphism
invariant  product is necessarily  sensitive on the differential
characteristics.

\section*{Acknowledgements}  This research was  supported in part
by Albert Einstein MPI, CGPG of Pennstate University, and  the Polish
Committee for Scientific Research under  grant no. 2  P03B 060 17.

\end{document}